\documentclass[aps,pra,preprint,groupedaddress,showpacs]{revtex4}

\usepackage{graphicx}

\begin{document}

\title{Laser cooling with a modified optical shaker}

\author{L. Marmet}
\email[]{louis..marmet@nrc.ca}
\affiliation{Institute for National Measurement Standards, National Research Council Canada, Ottawa, Canada K1A 0R6}

\date{Submitted to PRA 10 October 2008}

\begin{abstract}
  Some practical improvements are proposed for the ``optical-shaker'' laser-cooling technique [I.~S.~Averbukh and Y.~Prior, Phys.~Rev. Lett. \textbf{94}, 153002 (2005)].  The improved technique results in an increased cooling rate and decreases the minimum cooling temperature achievable with the optical shaker.  The modified shaker requires only one measurement of the force on the atoms before each cooling step, resulting in a simplification of the feedback electronics.  The force is inferred from the power variations of the transmitted laser beams and is used to determine the best moment at which the cooling steps are applied.  The temperature of the atomic system is automatically monitored, which allows maintaining an optimum cooling rate as the temperature decreases.  The improved shaker is simple to build, provides a faster rate of cooling, and can work in the microkelvin regime.  Numerical modeling shows a reduction by a factor of three in the required number of phase jumps and a lower temperature limit reduced by an order of magnitude compared to the initially proposed shaker.  The technique is also extendable to cooling in three dimensions.
\end{abstract}

\pacs{37.10.De}

\maketitle

\section{\label{sec:intro}Introduction}

  Laser cooling of atoms~\cite{metcalf.vanderstraten2003josab887} has made possible important experiments such as high-precision measurements of fundamental constants and the observation of Bose-Einstein condensation.  In many experiments, laser cooling relies on the nearly resonant exchange of momentum between the optical field and the atoms to obtain significant cooling.  However, conventional laser cooling is restricted to atomic species which offer an optical cycling transition as a nonconservative force.  Nonresonant interactions can address a larger variety of atoms and molecules by providing the means of trapping these particles and isolate them from strong resonant interactions with the lasers.  One disadvantage is that the conservative nature of the force does not naturally lend itself to cooling through an energy loss mechanism.  Only a few cooling techniques using nonresonant laser interactions have been proposed~\cite{ammann.chriestensen1997prl2088, vuletic.chu2000prl3787, averbukh.prior2005prl153002, vilensky.averbukh2007prl103002}.  Optical lattice clocks~\cite{katori.ovsiannikov2003prl173005, takamoto.katori2005n321, santra.ye2005prl173002, porsev.fortson2004pra021403} are good examples where weakly interacting laser beams isolate the trapped atoms from external perturbations.  Quantum computers also require cold atoms isolated from external influences to minimize decoherence~\cite{brennen.deutsch1999prl1060, garciaripoli.cirac2003prl127902}.

  One example of the nonresonant cooling technique is the optical shaker proposed by Averbukh and Prior~\cite{averbukh.prior2005prl153002} (AP shaker).  It combines the ideas of stochastic laser cooling~\cite{raizen.tajima1998pra4757} and Sisyphus cooling~\cite{dalibard.cohen-tannoudji1989josab2023} without the need for a resonant laser interaction.  Both mechanisms contribute to remove energy from trapped particles.  Stochastic cooling relies on the statistical nature of the spatial distribution of the particles in an optical potential to reduce their potential energy.  Such an optical potential can be produced with two counterpropagating laser beams.  For particles in a sinusoidal potential, the total potential energy of the particles $E$ is nonzero due to the random spatial distribution of the particles.  This energy $E$ is a sinusoidal function of the position of the optical potential.  By imposing a rapid phase jump on one laser beam at the appropriate time, the optical potential can be spatially displaced to decrease the potential energy of the particles without changing their kinetic energy.  The second mechanism, Sisyphus cooling, decreases the total kinetic energy of particles when they climb the potential hills.  The potential energy $E$ increases with time until another phase jump decreases it.  Repeated applications of phase jumps lowers the total energy of the system and reduces its temperature.  In stochastic cooling, the particles are not thermalized by collisions.  Throughout this paper, collisions are ignored and an equivalent temperature proportional to the total kinetic energy is used.

  The optical shaker produces phase jumps at the appropriate time and with the appropriate amplitude.  For the atoms to lose kinetic energy, a sufficient amount of time is given to let the system evolve so that the particles will have climbed the potential hills before the next phase jump.  A fixed time interval is chosen based on the wavelength of the optical potential and the average velocity of the particles.  The required size of the phase step is more difficult to evaluate.  As is explained below, the position of the optical potential must be inferred from the difference in power of the transmitted laser beams.  Before each jump, two consecutive measurements are required from which the size of the phase step is calculated numerically.  Fast electronics and a fast processor are required since under typical experimental conditions, only a few microseconds are available for the measurements and the calculation of a trigonometric function.  One drawback of the method is that it is limited to a minimum temperature in the several tens of microkelvins and the time constant to reach this temperature is of the order of several tens of seconds.  At this temperature, the power signals become very small and the cooling rate is limited by the shot noise on the detectors.

  The modified optical shaker proposed in the present paper leads to simplifications of the experimental system, an increase of the efficiency of the cooling process, and a lower cooling temperature limit.  To achieve this, the modified shaker monitors the time dependence of the transmitted powers.  The modified shaker is more efficient because it directly monitors the force on the particles instead of the potential energy.  A phase jump is triggered at the most appropriate time to achieve the fastest cooling.  Only one measurement is required before each jump, therefore reducing the required bandwidth of the electronic circuits.  Since the size of the phase step is a linear function of the time between the jumps, the need for a complex processor is eliminated.  With these improvements, atoms can be cooled down to temperatures near one microkelvin.  The theoretical basis describing the operation of the modified shaker is essentially the same as for the originally proposed AP shaker~\cite{averbukh.prior2005prl153002}.  This theory is presented in Sec.~\ref{sec:feedbackpotential} with a calculation of the cooling rate under optimized conditions.  The modified shaker is described in Sec.~\ref{sec:feedbackforce} with a theoretical derivation of it's efficiency.  In Sec.~\ref{sec:resultsdiscussion}, both a numerical and an analytical derivation of the temperature as a function of the cooling time are presented.  A comparison is made between the numerical model and the analytical model.  The limitations resulting from shot noise on the detectors and the minimum achievable temperature are discussed.

\section{\label{sec:feedbackpotential}Feedback minimizing the potential energy}

  A qualitative description of the AP shaker~\cite{averbukh.prior2005prl153002} can be given as follows.  Consider first the simple model of a single particle moving along two counterpropagating laser beams.  A dipole force on the particle arises from the momentum kick when a photon is taken from one beam and transferred to the other beam via stimulated emission~\cite{dalibard.cohen-tannoudji1985josab1707, martin.pritchard1988prl515}.  Since the force depends on a transfer of light from one beam to the other, it is monitored directly by measuring the power difference between the two beams.  The use of a periodic sinusoidal potential in the description of the mechanism is justified since the force is conservative.  To cool the particle, its kinetic energy must be reduced.  This can be achieved in two steps.  First, the potential energy of the particle is reduced with a rapid translation of the optical potential to a position where the particle will be at a minimum.  The second step consists in waiting a sufficient time to let the particle move away from a minimum and lose kinetic energy.  These steps can be repeated numerous times to reduce considerably the kinetic energy of the particle.  Consider now $N$ particles moving along the optical potential.  Because the particles have random positions, the total potential energy of these particles will not completely vanish.  The total potential energy will be sinusoidally dependent on the position of the optical potential.  As in the case of one particle, one can rapidly translate the optical potential to a position where the total potential energy is minimized.  The system will then evolve and the particles, as an ensemble, will lose some kinetic energy and therefore reduce their temperature.  The velocity is assumed to be a thermal distribution, but the spatial distribution of the particles is nearly uniform across one wavelength of the periodic potential.  The spatial distribution is essentially independent of time, except immediately after a phase jump when there is a larger density near the potential minima.

  Quantitatively, the AP shaker operates on an ensemble of $N$ particles interacting with two linearly polarized and counterpropagating laser beams having an electric field amplitude $E_0 \cos(k_l z - \omega_l t + \phi)$ and $E_0 \cos{(k_l z + \omega_l t)},$ where $k_l=2\pi /\lambda.$  The beams have the same polarization and the same frequency $\omega_l=2\pi c /\lambda$ with a phase difference $\phi$ at $z=0.$  The total potential energy of the $N$ particles at positions $Z_i(t)$ is written
\begin{equation}
	\label{eq:potential}
	U(\phi, t)=-U_0 u(\phi, t)/2,
\end{equation}
where the normalized potential energy is $u(\phi, t)=\sum^N_{i=1}\cos[z_i(t) + \phi],$ the normalized position is $z_i(t)=2\pi Z_i(t)/\Lambda,$ and $\Lambda=\lambda /2.$  The total force on these atoms is then
\begin{equation}
	\label{eq:force}
	F(\phi, t)=-\pi U_0 f(\phi, t)/\Lambda,
\end{equation}
where the normalized force $f(\phi, t)=\sum^N_{i=1}\sin[z_i(t)+\phi],$ $U_0=\alpha E_0^2,$ and $\alpha$ is the nonresonant polarizability.  In what follows, a red-detuned laser beam is considered so that $\alpha > 0.$  Stochastic cooling uses the random fluctuations in the potential energy of the system $u(\phi, t)$ to decrease the temperature of the system.  By imposing a rapid phase jump from $\phi$ to $\phi + \delta\phi,$ the total potential energy is decreased without changing the kinetic energy of the particles.  For a given $u(\phi, t)$ and $f(\phi, t),$ the required phase step to bring the potential energy $U(\phi + \delta\phi , t)$ to a minimum is~\cite{averbukh.prior2005prl153002}
\begin{equation}
	\label{eq:phasestep}
	\delta\phi=\arctan\left( -f(\phi, t), +u(\phi, t) \right),
\end{equation}
where the notation $\arctan\left(\rho\sin(\theta), \rho\cos(\theta) \right)=\theta$ (for any $\rho > 0$ and $-\pi < \theta \leq \pi$) is used to resolve the angle ambiguity encountered with the regular $\arctan\left(\sin(\theta)/\cos(\theta) \right)$ function.  A phase jump causes the system to lose an energy equal to
\begin{equation}
	\label{eq:energyloss}
	N\Delta E=\frac{U_0}{2}\left[ +\sqrt{u^2(\phi, t)+f^2(\phi, t)} - u(\phi, t) \right],
\end{equation}
where the positive root is used~\cite{averbukh.prior2005prl153002}.  The efficiency of the cooling technique is quantified by the average energy loss per particle at each jump, $\left\langle \Delta E \right\rangle.$  Since the system is allowed to evolve for a sufficiently long time before a phase jump, the steady-state statistical distributions of the two variables $u(\phi, t)$ and $f(\phi, t)$ are used for the calculation of $\left\langle \Delta E \right\rangle.$  Random fluctuations cause these two variables to have a normal distribution with a standard deviation
\begin{equation}
	\label{eq:standarddeviation}
	\sigma=\sqrt{N/2}.
\end{equation}
Their average values are more difficult to evaluate because the spatial distribution of the particles is not a uniform random spatial distribution.  Fast particles tend to accumulate near the potential maxima where they are slower, while slow particles are trapped at the bottom of the potential wells.  The result is a sinusoidally modulated distribution in phase with the optical potential.  In general, the expectation value $\left\langle U(\phi, t) \right\rangle$ is not zero.  However, since the force is in quadrature with the potential, it is also in quadrature with the spatial modulation of the particle distribution and averages to a null value
\begin{equation}
	\label{eq:averageforce}
	\left\langle F(\phi, t) \right\rangle=0.
\end{equation}

  To evaluate the average potential $\left\langle U(\phi, t) \right\rangle,$ the spatial distribution of the particles needs to be known.  Slower particles near the top of the potential hills spend more time in a region of high potential than faster particles near the bottom of the potential wells.  The time-dependent potential energy is therefore positive for longer time periods than it is negative.  As a result, the average potential energy~\cite{averbukh.prior2005prl153002} is
\begin{equation}
	\label{eq:averagepotentialspeed}
  \left\langle U(\phi, t) \right\rangle_S = rNU_0/2
\end{equation}
for $U_0 << k_BT,$ where $r = U_0/(4k_BT)$ is a parameter proportional to the cooling laser intensity.  This relation is valid for a specific velocity class and also for a Maxwell-Boltzmann speed distribution (hence the subscript $S$).  However, when the $v_z$ component of a Maxwell-Boltzmann velocity distribution is considered, one finds that a large fraction of the particles have a small velocity component $v_z$ and remain trapped in the potential wells.  These slow particles have a negative time-averaged potential energy if their total energy (kinetic and potential) is smaller than $E_k = (0.327\ldots) U_0.$  Particles with a total energy above $E_k$ have a positive average potential energy, but that energy is not high enough to compensate the negative contribution from the larger number of slower particles.  The time-averaged potential energy can be estimated for a thermal distribution of the velocity components in equilibrium with the standing-wave potential.  This requires an integral done over the potential energy $\epsilon(z)=-U_0 \cos(z)/2$ of a particle at location $z$ multiplied by the energy distribution function of the number of particles~\cite{sears.salinger1975book} at that location:
$$ \left\langle U(\phi, t) \right\rangle_V =
	 -\frac{N}{2\pi}\int^{2\pi}_0 \epsilon(z)\exp\left( \frac{-\epsilon(z)}{k_BT} \right)dz.$$
The average potential energy for a Maxwell-Boltzmann velocity distribution is
\begin{equation}
	\label{eq:averagepotentialvelocity}
	\left\langle U(\phi, t) \right\rangle_V = -rN U_0 /2
\end{equation}
for a small potential $U_0$ such that $r<<1/\sqrt{2N}.$  Note the negative value of the averaged potential compared to Eq.~(\ref{eq:averagepotentialspeed}).  This negative potential energy is a result of the maximum value at $v_z = 0$ of the Maxwell-Boltzmann velocity distribution.  For $r>>1/\sqrt{2N},$ the average potential energy reaches the limit $\left\langle U(\phi, t) \right\rangle = -NU_0/2.$  This corresponds to having all the particles at the bottom of the potential wells.  Averaging the energy loss given by Eq.~(\ref{eq:energyloss}) over the statistical distributions and the average values of $f(\phi, t)$ and $u(\phi, t)$ [from Eqs.~(\ref{eq:standarddeviation}), (\ref{eq:averageforce}) and (\ref{eq:averagepotentialvelocity})], the average energy loss per particle at each jump~\cite{averbukh.prior2005prl153002} is
\begin{equation}
	\label{eq:averageenergyloss}
	\left\langle \Delta E\right\rangle=k_BT/(2N)
\end{equation}
and is independent of $U_0$ as long as the condition $r>1/\sqrt{2N}$ is satisfied.  This result assumes a Maxwell-Boltzmann velocity distribution of the particles.  Collisions are ignored in this model.

\begin{figure}
	\includegraphics[width = 3.375in]{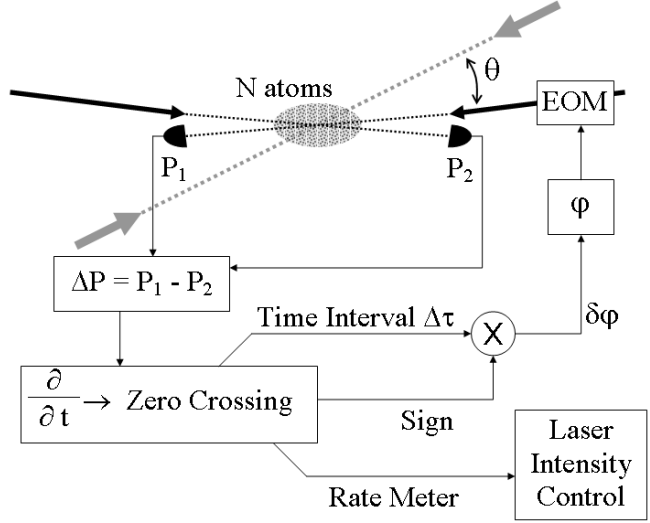}
	\caption{\label{fig:setup}
Schematic setup of the modified shaker.  The shaker beams are indicated by the long black arrows.  If separate trapping beams are used, they must be configured in a way that will not interfere with the shaker beams (gray arrows).  Two detectors measure the transmitted powers from which the power difference $\Delta P$ is calculated.  A zero-crossing detection circuit triggers the phase jump with an amplitude calculated from the slope of the derivative of $\Delta P$ and the time interval since the last phase jump.  The output of a rate meter controls the laser intensity for optimum cooling speed.
	}
\end{figure}

  Experimentally, cold particles are trapped by two nearly counterpropagating laser beams as shown in Fig.~\ref{fig:setup}.  The standing wave provides the potential wells used for the stochastic cooling.  An electro-optic modulator (EOM) can rapidly change the phase of one of the beams.  Two detectors measure the transmitted power of the two beams from which the force on the atoms $f(\phi, t)$ is calculated by monitoring the difference.  Conservation of momentum imposes that $\Delta P(\phi, t)=P_1-P_2=cF(\phi, t)=-\pi cU_0f(\phi, t)/\Lambda,$ where $c$ is the speed of light and $P_1$ and $P_2$ are the powers measured by the detectors.  An estimate of the stochastic power exchange using Eq.~(\ref{eq:standarddeviation}) gives
\begin{equation}
	\label{eq:powerexchange}
	\Delta P_{rms}=\frac{\pi cU_0}{\Lambda}\sqrt{\frac{N}{2}}.
\end{equation}
The AP shaker also needs the measurement of $u(\phi, t)$ to calculate the size of the necessary phase step with Eq.~(\ref{eq:phasestep}).  The potential energy is measured by briefly imposing a $\pi/2$ phase step and measuring $\Delta P(\phi+\pi/2,t)=cF(\phi+\pi/2,t)=-\pi cU_0u(\phi, t)/\Lambda.$  The phase step calculated from Eq.~(\ref{eq:phasestep}) is then added to the phase $\phi$ to increase the number of particles near the bottom of the potential wells.  According to the mechanism of Sisyphus cooling, the particles will climb the potential hills, gain potential energy, and lose kinetic energy.  Successive applications of the appropriate phase steps $\delta \phi$ will reduce the total energy of the system, therefore resulting in cooler particles.

  In order to give enough time for the system to evolve sufficiently before a phase jump, the time between jumps $t_j$ is chosen to be near the time of an oscillation period of low-energy particles trapped by the wave $t_\Omega=(2r)^{-1/2}t_\Lambda,$ where $t_\Lambda=\Lambda/v_{rms}$ is a characteristic time of the evolution of the system and $v_{rms} = \sqrt{k_BT/m}$ is the root-mean-square velocity of the particles at temperature $T$ in one dimension.  The laser intensities are chosen so that $r\approx1/\sqrt{2N}.$  This choice of parameters will result in an average energy loss given by Eq.~(\ref{eq:averageenergyloss}) and also produce reasonably sized phase steps $\delta\phi$ of the order of one radian.  A simpler feedback scheme~\cite{averbukh.prior2005prl153002} is possible where only a single measurement of the sign of $\Delta P(\phi, t)$ is needed.  A fixed phase step $\pm\delta\phi_0$ is applied to displace the standing wave opposite to the direction of the instantaneous force.  In this case, however, a phase jump will result in a smaller average energy loss per particle of $\left\langle \Delta E\right\rangle=k_BT/(\pi N).$  Adaptive lowering of $U_0$ is indicated to keep the optimum cooling conditions.

\begin{figure}
	\includegraphics[width = 3.375in]{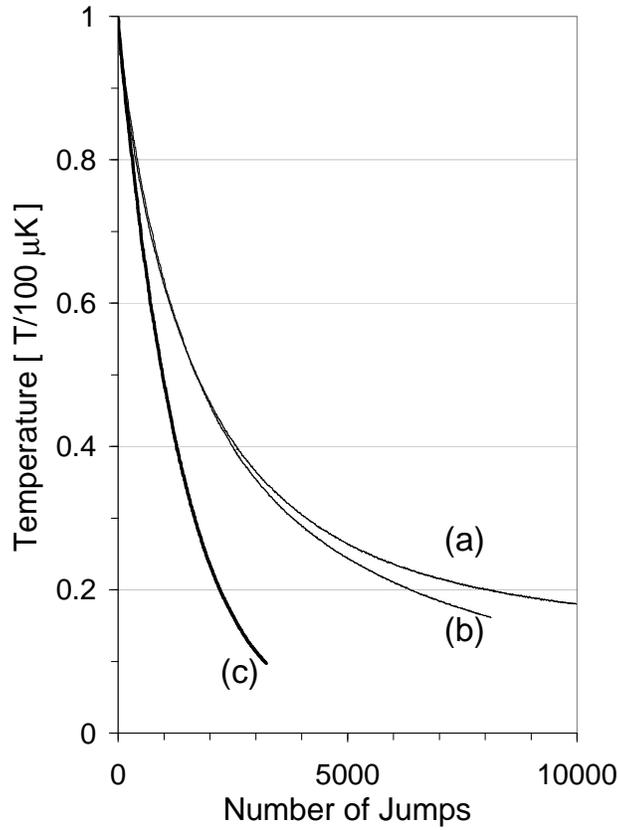}
	\caption{\label{fig:temperaturevsnumber}
Temperature as a function of the number of phase jumps.  In (a), the phase step of the AP shaker is determined from two measurements.  Curve (b) is the same technique with adaptive laser power and a shorter time between jumps.  Curve (c) shows the cooling using the method proposed in this paper.  The curves are the result of an average of fifteen runs.
	}
\end{figure}

\begin{figure}
	\includegraphics[width = 3.375in]{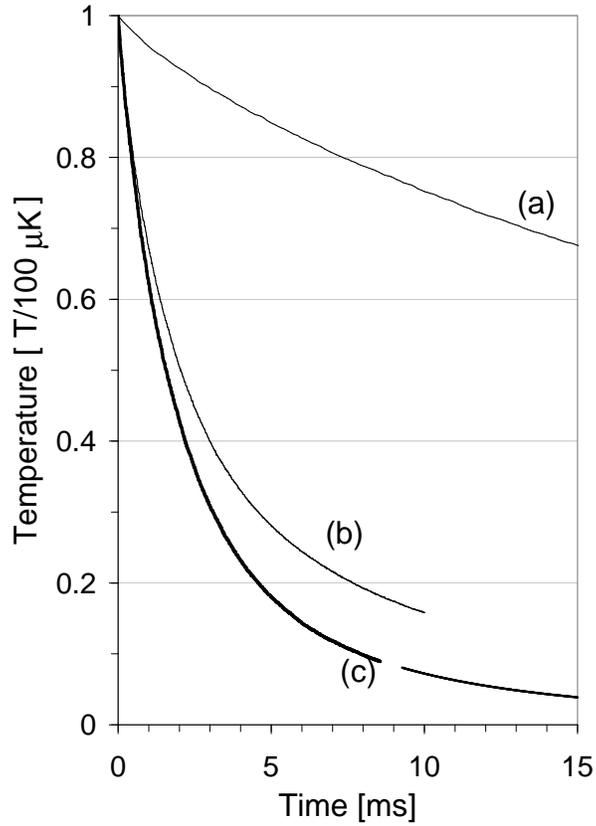}
	\caption{\label{fig:temperaturevstime}
Temperature as a function of time.  In (a), the phase step for the AP shaker is determined from two measurements.  The time between phase jumps is equal to the period of oscillation of low-energy particles trapped by the wave.  Curve (b) is the same technique with adaptive laser intensity optimized for the fastest cooling rate and a shorter time between jumps $t_j \approx 0.22 t_\Lambda.$  In (c), cooling is obtained using the modified shaker.  Because there is no need to let the system freely evolve and remix between two consecutive phase jumps, the cooling time is significantly shorter.  The temperature follows the expected theoretical curve shown as a thin line after $9$~ms.  The curves are the result of an average of fifteen runs.
	}
\end{figure}

  The results of the example given in~\cite{averbukh.prior2005prl153002} with $r=0.05$ and $t_j=t_\Omega$ are reproduced in Fig.~\ref{fig:temperaturevsnumber}(a), showing the temperature as a function of the number of phase jumps.  It is also interesting to know how quickly the cooling occurs by plotting the temperature as a function of time in Fig.~\ref{fig:temperaturevstime}(a).  An initial temperature of $100~\mu$K is used in this example.  One can ask how much time is needed between jumps in order to achieve the fastest cooling rate - that is, to maximize the energy loss per unit time and yet provide enough time for the system to evolve and mix sufficiently.  One could argue that some time after a jump, the system reaches a state of highest potential energy and lowest kinetic energy when most of the particles go from the bottom to the top of a potential hill.  The time required for this is about $t_j=t_\Lambda/2,$ which is much shorter than $t_\Omega.$  Numerical simulations show that the efficiency of the phase jumps remains constant for times as short as $t_j \approx 0.22t_\Lambda,$ but that the cooling rate is improved by a large factor.  As suggested in~\cite{averbukh.prior2005prl153002}, a decreasing laser power allows a more efficient cooling as the temperature decreases.  The numerical simulation was repeated with a decreasing laser power kept at a level proportional to the temperature and a faster jump rate $t_j.$  Optimization of the power gives the fastest cooling rate with an initial $r\approx0.07$ for $N=1000$ particles.  The results are plotted in Figs.~\ref{fig:temperaturevsnumber}(b) and ~\ref{fig:temperaturevstime}(b), showing a net improvement of the cooling efficiency as a function of time.  With a varying laser intensity, the cooling process follows Eq.~(\ref{eq:averageenergyloss}) over a wide range of temperatures.

\section{\label{sec:feedbackforce}Feedback minimizing the force}

  A simplification of the experimental setup resulting in a faster cooling rate is presented in this section.  To understand how the improved technique works, consider again the simple system with a single particle rolling up and down the potential hills.  When the particle reaches the top of the hill, the largest energy loss is obtained with a phase step $\delta\phi=\pi,$ which brings the particle to the bottom of a well.  As discussed above, it is difficult to know when the particle reaches that position since one can only directly measure the force on that particle $F(\phi, t),$ not the potential $U(\phi, t).$  A possible alternative is to wait until the particle reaches half the height of the potential well, then impose a phase step $\delta\phi=\pm\pi /2.$  This has the advantage that the force reaches a measurable extremum when it is time for a phase jump.  The sign of the phase step is readily determined from the sign of the force.  The process is different when there are several particles in the system since the potential energy does not oscillate in quadrature with the force, but instead is an independent random process.  However, the technique is still applicable with a small modification.
  
  For reasonably small phase steps $|\delta\phi|<1$ near a potential minimum, the relation $\left|f(\phi, t)\right| << \left|u(\phi, t)\right|$ is valid.  From Eqs.~(\ref{eq:potential}) and (\ref{eq:averagepotentialvelocity}) we get $\left\langle u(\phi, t)\right\rangle_V=rN.$  This implies that $u(\phi, t)$ is often positive.  In this case, Eq.~(\ref{eq:energyloss}) simplifies to
\begin{equation}
	\label{eq:energylossimproved}
	\Delta E\approx\frac{U_0}{4N}\frac{f^2(\phi, t)}{u(\phi, t)}.
\end{equation}
When a phase jump is applied, the largest energy loss is obtained with the largest possible value of $f^2(\phi, t)$ as in the one-particle example given above.  By monitoring the force and waiting until it reaches an extremum, the phase jump actually produces the largest energy loss.  Replacing $u(\phi, t)$ by its average value $rN$ and replacing $f^2(\phi, t)$ by the variance $N/2,$ Eq.~(\ref{eq:energylossimproved}) gives $\left\langle \Delta E\right\rangle \approx k_BT/(2N),$ the same energy loss obtained with the more complex method using measurements of the potential energy and the force on the particles given by Eq.~(\ref{eq:averageenergyloss}).  Because only one measurement is needed instead of two, the bandwidth of the detection system can be halved, resulting in an improved signal-to-noise ratio.  Additionally, the time between jumps is automatically controlled by the measurement of the force and automatically adapts itself to the dynamics of the system.

  One must determine the sign and magnitude of the phase step that will provide the best cooling.  After some time $\Delta\tau$ the force reaches an extremum and assuming that $u(\phi, t)$ is positive, the sign of the phase step is inferred from the sign of $f(\phi, t)$ and Eq.~(\ref{eq:phasestep}).  Since no direct measurement of the potential $u(\phi, t)$ is available, the magnitude of the required phase step cannot be calculated.  One could use a constant value for the phase step $\delta\phi,$ but this would reduce the efficiency of the cooling process and is not adaptable to different temperatures.  One could also assume a constant value of the potential, but an accurate measurement of $f(\phi, t)$ and a calculation of the arctangent is required.  An easier solution is to use the measured time $\Delta\tau$ elapsed since the last phase jump.  Just after a phase jump which brings the force to $f(\phi, t)\approx 0,$ the important point to notice is that the force evolves according to a random walk, which causes the value of $f^2(\phi, t)$ to increase linearly with time.  The required phase step also increases linearly with time as
\begin{equation}
	\label{eq:requiredphasestep}
	\delta\phi = \pm\omega(r,v_{rms})\Delta\tau.
\end{equation}
The equations of motion show that the time evolution of the system is inversely proportional to the velocity $v_{rms}.$  This suggests the use of a variable phase step factor $\omega(r, v_{rms})=\phi(r) v_{rms}/\Lambda,$ where the proportionality factor $\phi(r)$ is not explicitly dependent on $v_{rms}.$

\section{\label{sec:resultsdiscussion}Numerical results and discussion}

  Using numerical simulations, the fastest cooling is obtained with the optimized parameters $r\approx 0.098$ and $\phi(r)=0.85$ for $N=1000$ particles.  The good performance of the technique is readily visible in Figs.~\ref{fig:temperaturevsnumber}(c) and \ref{fig:temperaturevstime}(c).  A $40$\% reduction in cooling time is obtained compared to the optimum adaptive cooling rate obtained with the AP shaker~\cite{averbukh.prior2005prl153002}.  With optimum cooling conditions, the average size of a phase step $\left\langle \left| \delta\phi \right| \right\rangle \approx 0.21$ satisfies the above requirement of a small angle.  Since the jumps occur when $f^2(\phi, t)$ is at a maximum, a larger $\Delta E$ per jump is obtained.  The average energy loss per particle per jump becomes as large as
\begin{equation}
	\label{eq:energylossgood}
	\left\langle \Delta E\right\rangle \approx 1.4 k_BT/(2N).
\end{equation}
The $40$\% improvement over the estimated energy loss results from the fact that the variance of $f(\phi, t)$ is actually larger than $N/2$ at the time of the jump.  The larger variance shows that the system has not yet evolved to a random distribution just before the jump.  One sees the net advantage in choosing the appropriate time before imposing the phase jump as opposed to evenly distributing the jumps in time.  With a larger number of atoms, the optimum energy loss per jump given by Eq.~(\ref{eq:energylossgood}) is achieved by scaling down the power of the shaker beams as $1/\sqrt{N}.$  This maintains the average size of the phase steps near the optimum value of $0.21$~rad.

  Experimentally, the technique can be implemented in the following way.  In order to detect the extremum of the force on the atoms, a differentiating circuit is used to process the signal $\Delta P.$  This is indicated by $\partial/\partial t$ in Fig.~\ref{fig:setup}.  The circuit is easily built with a high-frequency pass {\itshape RC} filter.  The extremum occurs when $\partial \Delta P/\partial t = 0$ and is measured with a zero-crossing detection circuit.  Because of the increased high-frequency noise that could result at the output of the filter, a high-frequency cutoff filter should be added.  Numerical simulations showed that a filter with a response time $t_\Lambda$ is a good choice.  Three signals are then derived from the zero-crossing detection circuit: (a) a signal proportional to the time interval $\Delta\tau$ since the last zero crossing (i.e., a ramp generator), (b) the sign of the derivative just after the zero crossing (i.e., a comparator), and (c) a trigger signal for a rate meter.  When multiplied together, the first two signals give the value of $\delta\phi$ to be added to the phase of the EOM as per Eq.~(\ref{eq:requiredphasestep}).  The rate meter measures the average number of phase jumps per unit time interval and gives a signal proportional to the velocity $\nu_\Lambda = C_R v_{rms}.$  Numerical simulations show that $C_R = 4.2/\Lambda.$

  The equations of motion also show that to keep the system optimized, the parameter $r = U_0/(4k_BT)$ should be kept constant at the value determined numerically as $r \approx 3.1/\sqrt{N}.$  This imposes that for a given number of atoms, the potential $U_0$ remains proportional to the temperature as a function of time - that is, proportional to $\nu_\Lambda^2.$  This value can easily be obtained with a squaring circuit connected to the output of the rate meter.  This technique can easily be implemented as an all-analog circuit, therefore making it easy to achieve a fast response time of $100$~ns needed for initial temperatures of the order of $10$~mK.  It also has a net advantage over the method requiring two measurements and rapid digital processing for each phase jump~\cite{averbukh.prior2005prl153002, vilensky.prior2006pra063402}.  With this adaptive response implemented, the rate of the phase jumps is proportional to $v_{rms}$ or $\sqrt{T}$ and the energy loss follows Eq.~(\ref{eq:energylossgood}).  This is confirmed by the results plotted in Fig.~\ref{fig:temperaturevstime}(c), where the thin line extended after $9$~ms shows the temperature of the system calculated with the differential equation
$$\partial T(t)/\partial t \approx -1.4 C_R v_{rms} T(t)/N \propto -T(t)^{3/2}/N.$$

  The improved performance of the modified shaker is demonstrated here with a few examples.  A cooling time of the order of several seconds is reported for the original AP shaker~\cite{averbukh.prior2005prl153002} with the parameters $N=10^6$ cesium atoms initially at $15$~mK, a laser power $P=2~\mu W$ focused to a beam waist $w_0 = 10~\mu$m and $\lambda = 1~\mu$m with a detuning $\Delta = 100 \Gamma,$ where $\Gamma$ is the decay rate of the upper level of a two-level atom.  The cooling time is reduced to $300$~ms with the modified shaker method. (In this case, the optimum power is $0.12~\mu$W at $15$~mK, and the cooling time is defined as the time required to reach $e^{-2} \approx 13.5\%$ of the initial temperature.)  Another advantage of the improved technique is that it does not depend on an accurate measurement of the dc power of the transmitted laser beams (the AP shaker requires a power balance good to a few percent).  Instead, the modified shaker measures power fluctuations.  This is useful since at colder temperatures the stochastic power exchange $\Delta P$ becomes small and difficult to measure~\cite{morrow.raithel2002prl093003} as it scales proportionally to the intensity of the laser beams, which is adaptively decreased to maintain optimum cooling.  The use of a time differentiating circuit in the modified shaker improves performance since the output is insensitive to a possible dc signal which could appear, for example, as a result of interference from reflections on surfaces.

  The lowest temperature achievable with this technique is limited by the detector's shot noise and the intensity noise on the laser beams.  The effects of intensity noise have been discussed previously~\cite{vilensky.prior2006pra063402}.  The shot-noise limit for the modified shaker is reduced because the detectors operate with half the bandwidth of the AP shaker.  Thus, colder temperatures can be achieved.  The signal to be detected can be estimated~\cite{averbukh.prior2005prl153002} as $\Delta P/P \sim 4\sqrt{2N}\alpha/(\lambda\epsilon_0 w_0^2),$ where $\alpha \sim 4\pi \epsilon_0 (c/\omega_0)^3 (\Gamma/\Delta).$  For the best cooling efficiency, the conditions derived above require a shaker power $P \propto w_0^2 T (\Delta/\Gamma) /\sqrt{N}.$  This results in a detected signal $\Delta P \propto T.$  On the other hand, the shot-noise power is $P_N \propto \sqrt{P\eta B},$ where $B \propto \sqrt{T}$ is the required bandwidth and $\eta$ is the number of measurements needed for the shaker algorithm.  At the minimum cooling temperature, these two powers are equal and
$$T_{min} \approx 0.018~\mu \text{K}\ \frac{\text{w}_0^4 \eta^2}{N} \left(\frac{\Delta}{\Gamma}\right)^2,$$
where w$_0$ is the beam waist in $\mu$m and the constant is found from the optimum parameters calculated above.  In the example above for the AP shaker, $\eta = 2$ and the power difference becomes equal to the quantum noise at $T_{min} = 41~\mu$K.  The modified shaker requires only half of the bandwidth used by the AP shaker.  With $\eta = 1$ the minimum temperature is reduced to $T_{min} = 10~\mu$K.  The time required to cool from $74~\mu$K to $10~\mu$K is $4$~s.  These results are consistent with the conclusions that the minimum temperature obtainable with an optical shaker is not limited by the photon recoil limit, but by the square of the resolution of the measurement~\cite{ivanova.ivanov2005jetp482}.  If the coldest temperature is the goal to be achieved, a smaller detuning and a small beam waist should be considered.

  As the atoms are cooled, the optimum cooling power will decrease and the trapping potential will become shallower.  If the shaker beams are used as trapping beams, the trap depth might not be sufficient to hold the atoms against gravity.  However, the shaker beams can be distinct from the trapping beams.  The operation of the optical shaker is possible with separate trapping beams, as long as the standing waves produced by the beam pairs do not spatially coincide and that no power from the trapping beams reaches the power detectors.  The first condition is accomplished by using an angle between the two types of beams.  The second condition is achieved with a frequency offset to avoid producing a standing wave between the two types of beams.  The power in the shaker beams can be decreased independently of the power in the trapping beams, therefore enabling efficient cooling.  This configuration is depicted in Fig.~\ref{fig:setup}.  It also has the significant advantage that a lower cooling temperature is achievable.  If a small angle is used between the shaking and tapping beams, the atoms see a potential with an effective wavelength $\Lambda$ longer by a factor $\csc{(\theta)}$ in a direction perpendicular to the trapping beams.  The atoms have a larger amplitude of oscillation in that direction, which has the effect of decreasing the value of $C_R.$  The cooling speed is slower, but since the phase jump rate is decreased, a smaller bandwidth can be used for the detectors, therefore decreasing the minimum achievable temperature.  For example, an angle $\theta \approx 0.32 \approx 18.4^\circ$ decreases the cooling rate by a factor of $\sqrt{10},$ but reduces $T_{min}$ by a factor of ten.  In some applications the trapping beams have to be tuned very far away from any resonant level.  However, the cooling beams can be detuned by only a few GHz for the optimal cooling efficiency and the coldest temperature.

  The technique is easily generalized to an $n$-dimensional shaker built using a minimum of $n+1$ shaker beams~\cite{grynberg.salomon1993prl2249}.  In that case, phase control of $n$ beams is required based on $n+1$ power measurements.  In general, the phase control on beam number $i$ of an $n$-dimensional shaker uses the derivative of $\Delta P_i = \left( \sum _{j\neq i} P_j \right) - P_i.$  If trapping beams are present, they must allow some motion of the particles in at least one dimension.

\section{\label{sec:conclusions}Conclusions}

  A practical and efficient technique to implement optical shaking for laser cooling is described.  The technique combines the simplicity of the electronic feedback system and the adaptability required for efficient cooling over a large temperature range.  The faster response time of the feedback loop improves the limit imposed by its finite response~\cite{vilensky.prior2006pra063402}.  The modified shaker does not let the system go back to a random distribution between each shake, therefore improving the cooling speed compared to the AP shaker~\cite{averbukh.prior2005prl153002}.  The reason behind the improvement is that this technique is based on minimizing the force instead of the potential energy, which takes advantages of the correlation in the distribution of the atoms immediately after a cooling step.  By responding to immediate changes in the power fluctuations, it provides a high cooling rate that exceeds the rate obtained when predetermined time intervals are used between jumps.  This makes the operation of the improved shaker possible down to tens of microkelvins.  When used with additional trapping beams, lower temperatures can be achieved.  A fast cooling rate can become especially important in an optical lattice clock where the atoms need to be kept in the trapping potential at the coldest possible temperature for long periods of time.  The simplicity of this scheme makes it a promising configuration for a demonstration of the modified shaker cooling.

\bibliography{shaker}

\begin{thebibliography}{20}
\expandafter\ifx\csname natexlab\endcsname\relax\def\natexlab#1{#1}\fi
\expandafter\ifx\csname bibnamefont\endcsname\relax
  \def\bibnamefont#1{#1}\fi
\expandafter\ifx\csname bibfnamefont\endcsname\relax
  \def\bibfnamefont#1{#1}\fi
\expandafter\ifx\csname citenamefont\endcsname\relax
  \def\citenamefont#1{#1}\fi
\expandafter\ifx\csname url\endcsname\relax
  \def\url#1{\texttt{#1}}\fi
\expandafter\ifx\csname urlprefix\endcsname\relax\def\urlprefix{URL }\fi
\providecommand{\bibinfo}[2]{#2}
\providecommand{\eprint}[2][]{\url{#2}}

\bibitem[{\citenamefont{Metcalf and van~der
  Straten}(2003)}]{metcalf.vanderstraten2003josab887}
\bibinfo{author}{\bibfnamefont{H.~J.} \bibnamefont{Metcalf}} \bibnamefont{and}
  \bibinfo{author}{\bibfnamefont{P.}~\bibnamefont{van~der Straten}},
  \bibinfo{journal}{J. Opt. Soc. Am. B} \textbf{\bibinfo{volume}{20}},
  \bibinfo{pages}{887} (\bibinfo{year}{2003}).

\bibitem[{\citenamefont{Ammann and
  Christensen}(1997)}]{ammann.chriestensen1997prl2088}
\bibinfo{author}{\bibfnamefont{H.}~\bibnamefont{Ammann}} \bibnamefont{and}
  \bibinfo{author}{\bibfnamefont{N.}~\bibnamefont{Christensen}},
  \bibinfo{journal}{Phys. Rev. Lett.} \textbf{\bibinfo{volume}{78}},
  \bibinfo{pages}{2088} (\bibinfo{year}{1997}).

\bibitem[{\citenamefont{Vuleti\'{c} and Chu}(2000)}]{vuletic.chu2000prl3787}
\bibinfo{author}{\bibfnamefont{V.}~\bibnamefont{Vuleti\'{c}}} \bibnamefont{and}
  \bibinfo{author}{\bibfnamefont{S.}~\bibnamefont{Chu}},
  \bibinfo{journal}{Phys. Rev. Lett.} \textbf{\bibinfo{volume}{84}},
  \bibinfo{pages}{3787} (\bibinfo{year}{2000}).

\bibitem[{\citenamefont{Averbukh and
  Prior}(2005)}]{averbukh.prior2005prl153002}
\bibinfo{author}{\bibfnamefont{I.~S.} \bibnamefont{Averbukh}} \bibnamefont{and}
  \bibinfo{author}{\bibfnamefont{Y.}~\bibnamefont{Prior}},
  \bibinfo{journal}{Phys. Rev. Lett.} \textbf{\bibinfo{volume}{94}},
  \bibinfo{eid}{153002} (\bibinfo{year}{2005}).

\bibitem[{\citenamefont{Vilensky et~al.}(2007)\citenamefont{Vilensky, Prior,
  and Averbukh}}]{vilensky.averbukh2007prl103002}
\bibinfo{author}{\bibfnamefont{M.~Y.} \bibnamefont{Vilensky}},
  \bibinfo{author}{\bibfnamefont{Y.}~\bibnamefont{Prior}}, \bibnamefont{and}
  \bibinfo{author}{\bibfnamefont{I.~S.} \bibnamefont{Averbukh}},
  \bibinfo{journal}{Phys. Rev. Lett.} \textbf{\bibinfo{volume}{99}},
  \bibinfo{eid}{103002} (\bibinfo{year}{2007}), \bibinfo{note}{and references
  therein}.

\bibitem[{\citenamefont{{Katori} et~al.}(2003)\citenamefont{{Katori},
  {Takamoto}, { Pal'Chikov}, and
  {Ovsiannikov}}}]{katori.ovsiannikov2003prl173005}
\bibinfo{author}{\bibfnamefont{H.}~\bibnamefont{{Katori}}},
  \bibinfo{author}{\bibfnamefont{M.}~\bibnamefont{{Takamoto}}},
  \bibinfo{author}{\bibfnamefont{V.~G.} \bibnamefont{{ Pal'Chikov}}},
  \bibnamefont{and} \bibinfo{author}{\bibfnamefont{V.~D.}
  \bibnamefont{{Ovsiannikov}}}, \bibinfo{journal}{Phys. Rev. Lett.}
  \textbf{\bibinfo{volume}{91}}, \bibinfo{pages}{173005}
  (\bibinfo{year}{2003}).

\bibitem[{\citenamefont{Takamoto et~al.}(2005)\citenamefont{Takamoto, Hong,
  Higashi, and Katori}}]{takamoto.katori2005n321}
\bibinfo{author}{\bibfnamefont{M.}~\bibnamefont{Takamoto}},
  \bibinfo{author}{\bibfnamefont{F.-L.} \bibnamefont{Hong}},
  \bibinfo{author}{\bibfnamefont{R.}~\bibnamefont{Higashi}}, \bibnamefont{and}
  \bibinfo{author}{\bibfnamefont{H.}~\bibnamefont{Katori}},
  \bibinfo{journal}{Nature (London)} \textbf{\bibinfo{volume}{435}},
  \bibinfo{pages}{321} (\bibinfo{year}{2005}).

\bibitem[{\citenamefont{Santra et~al.}(2005)\citenamefont{Santra, Arimondo,
  Ido, Greene, and Ye}}]{santra.ye2005prl173002}
\bibinfo{author}{\bibfnamefont{R.}~\bibnamefont{Santra}},
  \bibinfo{author}{\bibfnamefont{E.}~\bibnamefont{Arimondo}},
  \bibinfo{author}{\bibfnamefont{T.}~\bibnamefont{Ido}},
  \bibinfo{author}{\bibfnamefont{C.~H.} \bibnamefont{Greene}},
  \bibnamefont{and} \bibinfo{author}{\bibfnamefont{J.}~\bibnamefont{Ye}},
  \bibinfo{journal}{Phys. Rev. Lett.} \textbf{\bibinfo{volume}{94}},
  \bibinfo{eid}{173002} (\bibinfo{year}{2005}).

\bibitem[{\citenamefont{Porsev et~al.}(2004)\citenamefont{Porsev, Derevianko,
  and Fortson}}]{porsev.fortson2004pra021403}
\bibinfo{author}{\bibfnamefont{S.~G.} \bibnamefont{Porsev}},
  \bibinfo{author}{\bibfnamefont{A.}~\bibnamefont{Derevianko}},
  \bibnamefont{and} \bibinfo{author}{\bibfnamefont{E.~N.}
  \bibnamefont{Fortson}}, \bibinfo{journal}{Phys. Rev. A}
  \textbf{\bibinfo{volume}{69}}, \bibinfo{pages}{021403(R)}
  (\bibinfo{year}{2004}).

\bibitem[{\citenamefont{Brennen et~al.}(1999)\citenamefont{Brennen, Caves,
  Jessen, and Deutsch}}]{brennen.deutsch1999prl1060}
\bibinfo{author}{\bibfnamefont{G.~K.} \bibnamefont{Brennen}},
  \bibinfo{author}{\bibfnamefont{C.~M.} \bibnamefont{Caves}},
  \bibinfo{author}{\bibfnamefont{P.~S.} \bibnamefont{Jessen}},
  \bibnamefont{and} \bibinfo{author}{\bibfnamefont{I.~H.}
  \bibnamefont{Deutsch}}, \bibinfo{journal}{Phys. Rev. Lett.}
  \textbf{\bibinfo{volume}{82}}, \bibinfo{pages}{1060} (\bibinfo{year}{1999}).

\bibitem[{\citenamefont{Garc\'ia-Ripoll and
  Cirac}(2003)}]{garciaripoli.cirac2003prl127902}
\bibinfo{author}{\bibfnamefont{J.~J.} \bibnamefont{Garc\'ia-Ripoll}}
  \bibnamefont{and} \bibinfo{author}{\bibfnamefont{J.~I.} \bibnamefont{Cirac}},
  \bibinfo{journal}{Phys. Rev. Lett.} \textbf{\bibinfo{volume}{90}},
  \bibinfo{pages}{127902} (\bibinfo{year}{2003}).

\bibitem[{\citenamefont{Raizen et~al.}(1998)\citenamefont{Raizen, Koga,
  Sundaram, Kishimoto, Takuma, and Tajima}}]{raizen.tajima1998pra4757}
\bibinfo{author}{\bibfnamefont{M.~G.} \bibnamefont{Raizen}},
  \bibinfo{author}{\bibfnamefont{J.}~\bibnamefont{Koga}},
  \bibinfo{author}{\bibfnamefont{B.}~\bibnamefont{Sundaram}},
  \bibinfo{author}{\bibfnamefont{Y.}~\bibnamefont{Kishimoto}},
  \bibinfo{author}{\bibfnamefont{H.}~\bibnamefont{Takuma}}, \bibnamefont{and}
  \bibinfo{author}{\bibfnamefont{T.}~\bibnamefont{Tajima}},
  \bibinfo{journal}{Phys. Rev. A} \textbf{\bibinfo{volume}{58}},
  \bibinfo{pages}{4757} (\bibinfo{year}{1998}).

\bibitem[{\citenamefont{Dalibard and
  Cohen-Tannoudji}(1989)}]{dalibard.cohen-tannoudji1989josab2023}
\bibinfo{author}{\bibfnamefont{J.}~\bibnamefont{Dalibard}} \bibnamefont{and}
  \bibinfo{author}{\bibfnamefont{C.}~\bibnamefont{Cohen-Tannoudji}},
  \bibinfo{journal}{J. Opt. Soc. Am. B} \textbf{\bibinfo{volume}{6}},
  \bibinfo{pages}{2023} (\bibinfo{year}{1989}).

\bibitem[{\citenamefont{Dalibard and
  Cohen-Tannoudji}(1985)}]{dalibard.cohen-tannoudji1985josab1707}
\bibinfo{author}{\bibfnamefont{J.}~\bibnamefont{Dalibard}} \bibnamefont{and}
  \bibinfo{author}{\bibfnamefont{C.}~\bibnamefont{Cohen-Tannoudji}},
  \bibinfo{journal}{J. Opt. Soc. Am. B} \textbf{\bibinfo{volume}{2}},
  \bibinfo{pages}{1707} (\bibinfo{year}{1985}).

\bibitem[{\citenamefont{Martin et~al.}(1988)\citenamefont{Martin, Oldaker,
  Miklich, and Pritchard}}]{martin.pritchard1988prl515}
\bibinfo{author}{\bibfnamefont{P.~J.} \bibnamefont{Martin}},
  \bibinfo{author}{\bibfnamefont{B.~G.} \bibnamefont{Oldaker}},
  \bibinfo{author}{\bibfnamefont{A.~H.} \bibnamefont{Miklich}},
  \bibnamefont{and} \bibinfo{author}{\bibfnamefont{D.~E.}
  \bibnamefont{Pritchard}}, \bibinfo{journal}{Phys. Rev. Lett.}
  \textbf{\bibinfo{volume}{60}}, \bibinfo{pages}{515} (\bibinfo{year}{1988}).

\bibitem[{\citenamefont{Sears and Salinger}(1975)}]{sears.salinger1975book}
\bibinfo{author}{\bibfnamefont{F.~W.} \bibnamefont{Sears}} \bibnamefont{and}
  \bibinfo{author}{\bibfnamefont{G.~L.} \bibnamefont{Salinger}},
  \emph{\bibinfo{title}{Thermodynamics, Kinetic Theory, and Statistical
  Thermodynamics}}, Addison-Wesley Principles of Physics
  (\bibinfo{publisher}{Addison-Wesley. Reading, MA}, \bibinfo{year}{1975}),
  \bibinfo{edition}{3rd} ed.

\bibitem[{\citenamefont{Vilensky et~al.}(2006)\citenamefont{Vilensky, Averbukh,
  and Prior}}]{vilensky.prior2006pra063402}
\bibinfo{author}{\bibfnamefont{M.~Y.} \bibnamefont{Vilensky}},
  \bibinfo{author}{\bibfnamefont{I.~S.} \bibnamefont{Averbukh}},
  \bibnamefont{and} \bibinfo{author}{\bibfnamefont{Y.}~\bibnamefont{Prior}},
  \bibinfo{journal}{Phys. Rev. A} \textbf{\bibinfo{volume}{73}},
  \bibinfo{eid}{063402} (\bibinfo{year}{2006}).

\bibitem[{\citenamefont{Morrow et~al.}(2002)\citenamefont{Morrow, Dutta, and
  Raithel}}]{morrow.raithel2002prl093003}
\bibinfo{author}{\bibfnamefont{N.~V.} \bibnamefont{Morrow}},
  \bibinfo{author}{\bibfnamefont{S.~K.} \bibnamefont{Dutta}}, \bibnamefont{and}
  \bibinfo{author}{\bibfnamefont{G.}~\bibnamefont{Raithel}},
  \bibinfo{journal}{Phys. Rev. Lett.} \textbf{\bibinfo{volume}{88}},
  \bibinfo{pages}{093003} (\bibinfo{year}{2002}).

\bibitem[{\citenamefont{Ivanova and Ivanov}(2005)}]{ivanova.ivanov2005jetp482}
\bibinfo{author}{\bibfnamefont{T.~Y.} \bibnamefont{Ivanova}} \bibnamefont{and}
  \bibinfo{author}{\bibfnamefont{D.~A.} \bibnamefont{Ivanov}},
  \bibinfo{journal}{JETP Lett.} \textbf{\bibinfo{volume}{82}},
  \bibinfo{pages}{482} (\bibinfo{year}{2005}).

\bibitem[{\citenamefont{Grynberg et~al.}(1993)\citenamefont{Grynberg, Lounis,
  Verkerk, Courtois, and Salomon}}]{grynberg.salomon1993prl2249}
\bibinfo{author}{\bibfnamefont{G.}~\bibnamefont{Grynberg}},
  \bibinfo{author}{\bibfnamefont{B.}~\bibnamefont{Lounis}},
  \bibinfo{author}{\bibfnamefont{P.}~\bibnamefont{Verkerk}},
  \bibinfo{author}{\bibfnamefont{J.-Y.} \bibnamefont{Courtois}},
  \bibnamefont{and} \bibinfo{author}{\bibfnamefont{C.}~\bibnamefont{Salomon}},
  \bibinfo{journal}{Phys. Rev. Lett.} \textbf{\bibinfo{volume}{70}},
  \bibinfo{pages}{2249} (\bibinfo{year}{1993}).

\end{thebibliography}

\end{document}